\documentclass[useAMS,usenatbib]{mn2e}

\usepackage{graphicx,psfrag}

\def \rgs {H$z$RGs}

\title[What are Protoclusters?]{What are Protoclusters? -- Defining High Redshift Galaxy Clusters and Protoclusters}
\author[Muldrew, Hatch \& Cooke]{Stuart I. Muldrew$^1$\thanks{E-mail:
stuart.muldrew@leicester.ac.uk}, Nina A. Hatch$^2$ and Elizabeth A. Cooke$^2$\\
$^1$Department of Physics and Astronomy, University of Leicester, University Road, Leicester, LE1 7RH, UK\\
$^2$School of Physics and Astronomy, University of Nottingham, Nottingham, NG7 2RD, UK}
\begin{document}

\date{Accepted 2015 June 27.  Received 2015 June 16; in original form 2015 March 16.}

\pagerange{1--14} \pubyear{2015}

\maketitle

\label{firstpage}

\begin{abstract}
We explore the structures of protoclusters and their relationship with high redshift clusters using the Millennium Simulation combined with a semi-analytic model.  We find that protoclusters are very extended, with 90 per cent of their mass spread across $\sim35\,h^{-1}{\rm Mpc}$ comoving at $z=2$ ($\sim30\, \rm{arcmin}$). The `main halo', which can manifest as a high redshift cluster or group, is only a minor feature of the protocluster, containing less than 20 per cent of all protocluster galaxies at $z=2$. Furthermore, many protoclusters do not contain a main halo that is massive enough to be identified as a high redshift cluster.  Protoclusters exist in a range of evolutionary states at high redshift, independent of the mass they will evolve to at $z=0$. We show that the evolutionary state of a protocluster can be approximated by the mass ratio of the first and second most massive haloes within the protocluster, and the $z=0$ mass of a protocluster can be estimated to within 0.2 dex accuracy if both the mass of the main halo and the evolutionary state is known.  We also investigate the biases introduced by only observing star-forming protocluster members within small fields. The star formation rate required for line-emitting galaxies to be detected is typically high, which leads to the artificial loss of low mass galaxies from the protocluster sample.  This effect is stronger for observations of the centre of the protocluster, where the quenched galaxy fraction is higher.  This loss of low mass galaxies, relative to the field, distorts the size of the galaxy overdensity, which in turn can contribute to errors in predicting the $z=0$ evolved mass.  
\end{abstract}

\begin{keywords}
methods: numerical -- methods: statistical -- galaxies: clusters: general -- galaxies: formation -- galaxies: evolution -- cosmology: theory
\end{keywords}

\section{Introduction}
\label{intro}

In a cold dark matter universe with a cosmological constant ($\Lambda$CDM), structure forms through hierarchical growth with smaller haloes merging to form larger ones.  Galaxy clusters in the present day Universe are the most massive structures to have formed and were the result of the merging of many smaller haloes.  Clusters, typically, are virialised dark matter haloes of mass greater than $10^{14}\,{\rm M_{\odot}}$ containing a hot X-ray Intra-Cluster Medium (ICM) and red, passive galaxies.  

At higher redshift, $z>1.5$, most clusters were not the massive virialised haloes that we see today.  Instead we see their progenitors, a diffuse collection of haloes that will merge to make the final halo.  The term `protocluster' is often used to describe this state, but differing definitions of what a protocluster is exist in the literature.  While some define a protocluster as all the haloes at a given redshift that will merge to make the final cluster, others define it as being just the most massive progenitor halo, sometimes referred to as the main halo.  While using the latter definition dramatically reduces the observational expense, it risks missing galaxies undergoing environmental preprocessing and only captures part of what is going on in the forming cluster.  

Several high redshift galaxy clusters have now been detected through X-ray emission, the Sunyaev-Zel'dovich (SZ) effect, as well as through photometric redshift hunts in large deep surveys \citep{Gobat11,Stanford12,Zeimann12,Fassbender14,Andreon14}.  The properties of the ICM and galaxies indicate that these structures are already collapsed, i.e. these objects are single collapsed main haloes.  However, a great deal of cluster growth occurs at relatively late times \citep[$z<1$;][]{Chiang13}, and many of the galaxies and dark matter that end up in the $z=0$ cluster, will not be located in the main halo of the protocluster at high redshift.  In this paper we investigate how much of the matter and galaxies reside in the main halo compared to the entire protocluster as a function of redshift, and use this to investigate its significance.  Additionally we look at whether protoclusters with evolved main haloes are representative of all protoclusters, or are a subsample that are easier to detect.

Identifying protoclusters has so far been challenging due to their low number density and the faintness of distant galaxies.  One of the most successful methods for detecting protoclusters is to use High Redshift Radio Galaxies (\rgs) as a tracer population to locate overdense regions \citep[e.g.][]{LeFevre96,Pentericci00,Best03,Venemans07,Galametz10,Hatch11,Galametz13,Wylezalek13,Cooke14}.  These galaxies are among the most massive galaxies at all epochs \citep{Seymour07}, but the large galaxy overdensities that surround these radio-loud galaxies exceed that of similar mass radio-quiet galaxies \citep{Hatch14}. Both \citet{RamosAlmeida13} and \citet{Hatch14} concluded that dense environments foster the formation of radio-loud jets from AGN, which explains why \rgs~are excellent beacons of galaxy protoclusters and high redshift clusters.

An alternative technique is to identify protoclusters in large surveys with accurate photometric redshifts. Precise photometric redshifts at $z\ge2$ are difficult to obtain due to the Balmer break shifting into the near-infrared wavelength, but \citet{Spitler12} has shown that Virgo-like cluster progenitors can be found if medium-band near-infrared filters are used.  \citet{Chiang13} advocates a protocluster detection method that finds galaxy overdensities in $15\,{\rm Mpc}$ comoving windows; applying this method to the 1.62deg$^{2}$ COSMOS/UltraVISTA field \citep{Muzzin13} has resulted in 36 candidate structures \citep{Chiang14}. This method is effective because of the correlation between aperture density and halo mass \citep{Haas12,Muldrew12}, however at high redshift there is considerable uncertainty in this relation due to projection effects \citep{Shattow13}.

A number of techniques have been used to isolate the galaxies within high redshift clusters and protoclusters for further study.  Photometric redshifts from deep multi-band data can reach accuracies of $\Delta z/(1+z)=0.03$, and although (proto-)cluster galaxies have been selected using photometric redshifts \citep[e.g.][]{Tanaka10}, the sample is often incomplete and greatly contaminated by foreground and background galaxies.  One of the most successful techniques for locating clean samples of protocluster galaxies is using narrow-band filters with specific central wavelengths matched to the wavelength of an emission line from protocluster galaxies \citep[e.g.][]{Kurk04,Venemans07,Hatch11b,Cooke14}. The ideal line is H$\alpha$ since it is a strong line which is least affected by dust absorption. Selecting galaxies based on their line emission means only active galaxies are located, i.e. star-forming galaxies and AGN. This can limit our view of the protocluster in unexpected ways. Here we explore how this selection method can give a biased view of the protocluster.

In this paper we explore the galaxies that make up protoclusters using a semi-analytic model built upon the Millennium Simulation.  In Section \ref{sec:meth} we describe the simulations used and how we constructed the protocluster catalogue.  Using this mock catalogue, in Section \ref{sec:res}, we give an overview of the spatial properties of protoclusters and their member galaxies.  We then examine two fundamental issues concerning protoclusters: the relationship of the main progenitor halo (which is sometimes observed as the high redshift cluster) to the rest of the protocluster and the cluster's $z=0$ mass; and how our understanding of protoclusters is biased when only active protocluster galaxies are observed. In Section \ref{sec:sum} we summarise our findings and reflect on the implications they have for interpreting observations of protoclusters and high redshift clusters.

\section{Methods}
\label{sec:meth}

To construct a statistically large sample of galaxy clusters, whose evolution can be tracked back to high redshift, we used the \citet{Guo11} semi-analytic model applied to the Millennium Simulation \citep{Springel05}.  Clusters were identified as haloes with masses greater than $10^{14}\,h^{-1}{\rm M_{\odot}}$ at $z=0$, while protoclusters were defined as the cluster progenitors.

\subsection{The Millennium Simulation and Semi-Analytic Model}

\begin{figure*}
\begin{center}
 \psfrag{a}[][][1][0]{$x/(h^{-1}{\rm Mpc})$}
 \psfrag{b}[][][1][0]{$y/(h^{-1}{\rm Mpc})$}
 \psfrag{c}[l][][1][0]{$z=2$}
 \psfrag{d}[l][][1][0]{$z=1$}
 \psfrag{e}[l][][1][0]{$z=0$}
 \psfrag{f}[l][][1][0]{$M_{200}^{z=0}=10^{15.4}\,h^{-1}{\rm M_{\odot}}$}
 \psfrag{g}[l][][1][0]{$M_{200}^{z=0}=10^{14.8}\,h^{-1}{\rm M_{\odot}}$}
 \psfrag{h}[l][][1][0]{$M_{200}^{z=0}=10^{14.0}\,h^{-1}{\rm M_{\odot}}$}
 \includegraphics[width=150mm]{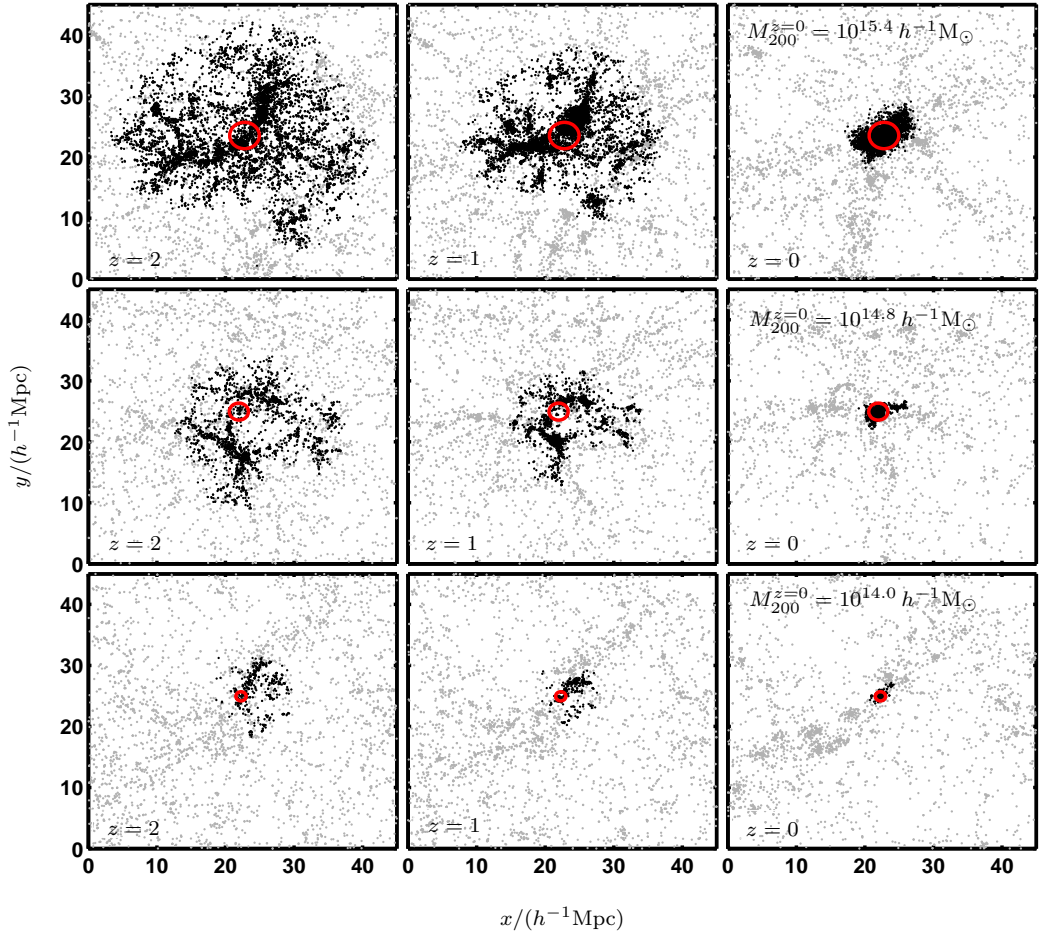}
  \caption{The spatial extent of protoclusters at $z=2$ (left panel), 1 (centre panel) and 0 (right panel), with final cluster masses of $M_{200}^{z=0}=10^{15.4}\,h^{-1}{\rm M_{\odot}}$ (top row), $10^{14.8}\,h^{-1}{\rm M_{\odot}}$ (middle row) and $10^{14.0}\,h^{-1}{\rm M_{\odot}}$ (bottom row).  Each window is $45\times45\,h^{-1}{\rm Mpc}$ comoving, which corresponds to 41 arcmin and 65 arcmin at $z=2$ and $z=1$ respectively \citep{Wright06}.  Black points represent a galaxy of stellar mass greater than $10^{8}\,h^{-1}{\rm M_{\odot}}$ that will end up in the cluster while grey points represent those that will not.  (Only 25 per cent of the background galaxies, grey points, are plotted to reduce image size.)  The red circle corresponds to the $z=0$ centre and comoving viral radius of the cluster.}
\label{fig:ext}
\end{center}
\end{figure*}

\begin{figure*}
\begin{center}
  \psfrag{a}[][][1][0]{$z$}
  \psfrag{b}[][][1][0]{$r_{\rm comoving}/(h^{-1}{\rm Mpc)}$}
  \psfrag{c}[l][][0.8][0]{$1\le M_{200}^{z=0}<4\times10^{14}\,h^{-1}{\rm M_{\odot}}$}
  \psfrag{d}[l][][0.8][0]{$4\le M_{200}^{z=0}<10\times10^{14}\,h^{-1}{\rm M_{\odot}}$}
  \psfrag{e}[l][][0.8][0]{$M_{200}^{z=0}\ge 10^{15}\,h^{-1}{\rm M_{\odot}}$}
    \psfrag{f}[][][1][0]{$r_{\rm physical}/(h^{-1}{\rm Mpc)}$}
      \psfrag{g}[][][1][0]{$r_{\rm angular}/{\rm arcmin}$}
  \includegraphics[width=180mm]{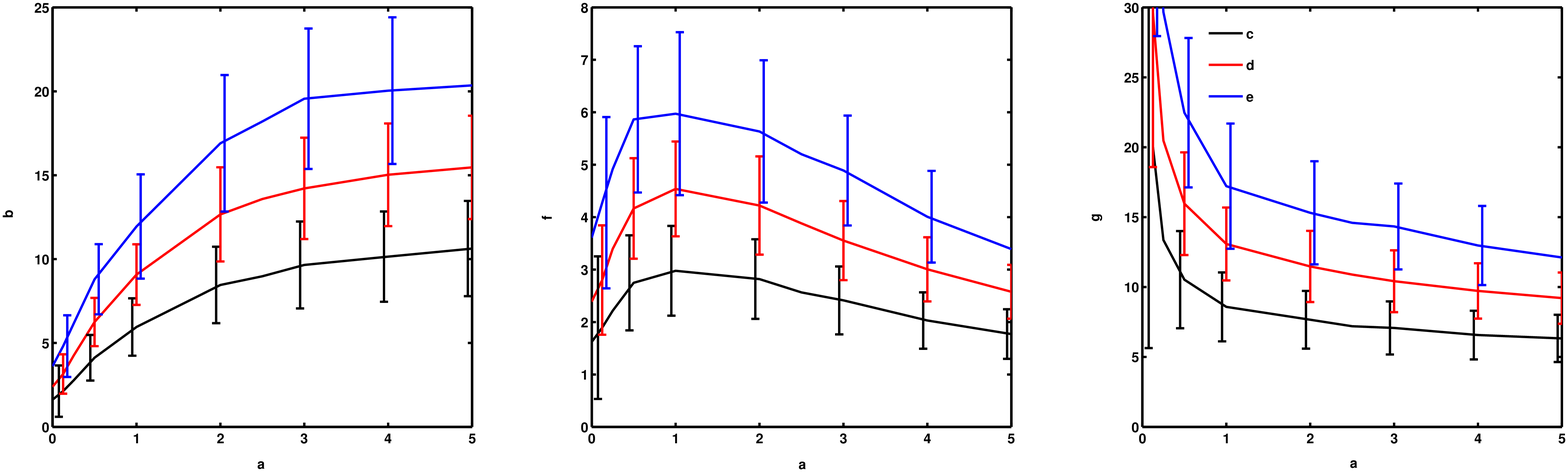}
  \caption{The average radius that encloses 90 per cent of the stellar mass of a protocluster at different redshifts, for binned $z=0$ cluster masses. The left panel represents comoving radius, centre panel the physical radius and right panel the angular projection.  Error bars represent $1\,\sigma$ scatter and are offset about the middle mass bin by $\delta z=0.05$ for clarity.  This radius is tightly correlated with the radius enclosing 90 per cent of the dark matter mass.}
  \label{fig:rad}
  \end{center}
\end{figure*}

The Millennium Simulation follows the evolution of $2160^3$ dark matter particles in a cube of comoving side length $500\,h^{-1}{\rm Mpc}$, using the $N$-body code \textsc{gadget}-2 \citep{Springel05a}.  It adopts a $\Lambda$CDM cosmology with parameters $\Omega_0=0.25$, $\Omega_\Lambda=0.75$, $h=0.73$, $n=1$ and $\sigma_8=0.9$ consistent with the Two-Degree Field Galaxy Redshift Survey \citep[2dFGRS;][]{Colless01} and the first-year \textit{Wilkinson Microwave Anisotropy Probe} data \citep[\textit{WMAP}-1;][]{Spergel03}.

Haloes were detected using a two-step procedure.  Firstly, a Friends-of-Friends algorithm \citep[FoF;][]{Davis85} with linking length, $b=0.2$, was used to identify haloes and these were then post-processed using \textsc{subfind} \citep{Springel01}.  All haloes with greater than 20 particles were used to construct merger trees.  We note similar results are found with other halo finders \citep{Muldrew11,Knebe11}.

To populate the simulation with galaxies, the \citet{Guo11} semi-analytic model was applied to the resulting merger trees.  This model is an updated version of that previously presented in \citet{Croton06} and \citet{DeLucia07} and gives a better fit to the redshift evolution in the galaxy stellar mass function.  The model includes prescriptions for gas infall, shock heating, cooling, star formation, stellar evolution, supernova feedback, black hole growth and feedback, metal enrichment, mergers, and tidal and ram-pressure stripping.  Full details of these implementations can be found within the previously referenced papers.  For the purpose of this study we cut the semi-analytic catalogue to only include galaxies with stellar masses greater than $10^8\,h^{-1}{\rm M_{\odot}}$.  This is above the resolution limit adopted by \citet{Guo11}, but is still below the detection threshold of most observational protocluster studies \citep[e.g.][]{Cooke14}.  All results that are dependent on stellar mass in this paper are presented against mass or with different minimum cuts to illustrate the effect of having a minimum galaxy mass cut.

The cosmological parameters used for the Millennium Simulation were in agreement with the results of \textit{WMAP}-1, but have become slightly discrepant with the latest values from \textit{Planck} \citep{Planck14}.  \citet{Angulo10} proposed a method of rescaling dark matter simulations to different cosmologies by reassigning the mass and position of particles and the redshift of the snapshot.  \citet{Guo13} applied this method to the Millennium Simulation to obtain a galaxy catalogue for \textit{WMAP}-7 cosmology \citep{Komatsu11}.  They found that the increased matter density, $\Omega_m$ offsets the effect of a decreased linear fluctuation amplitude, $\sigma_8$, which leads to very similar results for $z<3$.  This should have even less of an effect for \textit{Planck} cosmology, where the rescaling from \textit{WMAP}-1 is not as large \citep{Henriques14}.  Further comparison for protoclusters between \textit{WMAP}-1 and \textit{WMAP}-7 cosmology was made by \citet{Chiang13}, who found little difference in results.  This confirms that using a simulation based on \textit{WMAP}-1 cosmology will have little overall impact on our results.

\subsection{Protocluster Identification}

We identified galaxy clusters in the simulation exclusively on dark matter halo mass.  All haloes with $M_{200}\geq10^{14}\,h^{-1}{\rm M_{\odot}}$  at $z=0$, where $M_{200}$ is the mass enclosed by a sphere whose density is 200 times the critical density of the Universe, were defined as galaxy clusters.  This gave a total of $1,938$ clusters in our sample.  All semi-analytic galaxies that are members of the FoF haloes are then classed as galaxy cluster members.  Each halo consists of a `central' galaxy, which is at the centre of the halo, and `satellite' galaxies. 

For protoclusters, we trace the merger tree back in time to each redshift of interest.  For each $z=0$ cluster, we identify all the haloes at a given redshift that will merge to form it and identify this as the protocluster.  All galaxies that are associated with these haloes are then classed as protocluster members.  For example, at $z=2$ there are $1,938$ protoclusters, which are the progenitors of the $z=0$ clusters, but these are made up of $639,253$ individual haloes with a central galaxy of at least $M_{*}=10^8\,h^{-1}{\rm M_{\odot}}$.  Unless stated otherwise, the protocluster studies presented in this paper are for $z=2$.

\section{Results}
\label{sec:res}

The results are presented in three sections looking at different aspects of protoclusters.  Firstly in Section \ref{sec:dist} we explore the distribution of protocluster member galaxies to explain what protoclusters are.  In Section \ref{sec:MainHalo} we examine the relationship between protoclusters and their main haloes. Finally, in Section \ref{sec:LowMass} we explore how limiting our observations to only the active subset of protocluster galaxies can affect our understanding of protoclusters.

\subsection{The Distribution of Protocluster Galaxies}
\label{sec:dist}

To begin, we explore the distribution of galaxies that make up a protocluster.  Figure \ref{fig:ext} displays the spatial extent of protocluster galaxy members, with mass $M_{*}\ge10^{8}\,h^{-1}{\rm M_{\odot}}$ at $z=2$, 1 and 0 for $M_{200}^{z=0}$ masses of $10^{14.0}$, $10^{14.8}$ and $10^{15.4}\,h^{-1}{\rm M_{\odot}}$.  The red circle corresponds to the comoving $z=0$ virial radius of the cluster.  At $z=2$ the $M_{200}^{z=0}=10^{15.4}\,h^{-1}{\rm M_{\odot}}$ cluster (top left panel) extends to $45\,h^{-1}{\rm Mpc}$ comoving ($15\,h^{-1}{\rm Mpc}$ physical), showing a rich structure of haloes and filaments.  The structure is far from the collapsed single halo it becomes at $z=0$ (top right panel).  For the lower mass clusters, a similar filamentary distribution is visible at high redshift, but the overall spread is much smaller.

Due to the limit of instrumental fields-of-view, targeted observational imaging studies of protoclusters have typical windows of a few arcmin on a side (e.g.~$2.5\,{\rm arcmin}$ in \citealt{Cooke14} or $7\,{\rm arcmin}$ in \citealt{Koyama13}).  For $z=2$, and the cosmology of the Millennium Simulation, this corresponds to $2.8\,h^{-1}{\rm Mpc}$ and $7.7\,h^{-1}{\rm Mpc}$ comoving respectively \citep[determined using `The Cosmology Calculator';][]{Wright06}.  Comparing these to the full distribution of the protocluster, the left hand panels of Figure \ref{fig:ext}, demonstrates that in all but the lowest mass case, only a small area of the protocluster is being captured.  In the top panel, for the most massive cluster, the red circle corresponds to the $z=0$ virial radius of $2.16\,h^{-1}{\rm Mpc}$ comoving.  This circle would enclose the smaller aperture of \citet{Cooke14}.  This means that any observations of protoclusters carried out in this way are not following the entire protocluster, but are focussed on just the growth of the central region. 

To further illustrate the large spatial extent of the protocluster, we plot the radius that encloses 90 per cent of the stellar mass at different redshifts in Figure \ref{fig:rad}.  The 90 per cent stellar mass radius is strongly correlated to the 90 per cent dark matter mass radius making it an excellent measure of cosmological growth.  The protoclusters are binned by their $z=0$ mass and sizes are presented in comoving, physical and angular scale.  We explore the difference between defining cluster members using the FoF halo or virial radius in Appendix \ref{ap}.

In the comoving reference frame, protoclusters continually collapse, albeit gradually above $z=3$. At these high redshifts the protoclusters display a similar comoving size, for fixed $z=0$ mass, indicative of the shorter amount of cosmic time that passes compared to lower redshift and the late collapse of clusters.  The difference in size with mass at fixed redshift is also much larger at high redshift compared with the present day.  The large sizes are in agreement with those found by \citet{Chiang13} using an alternative measure of the protocluster's radius.

In physical units the behaviour of the protocluster appears quite different.  At high redshift protoclusters are still expanding with the Universe before collapse occurs after $z=1$. Therefore the global density of protoclusters decreases with time until $z\sim1$, after which they rapidly collapse. From an observational point of view, protoclusters extend over approximately the same angle across the sky from $z\sim5$ to $z\sim1$. This means that protocluster detection algorithms do not need to search over different sized apertures to locate protoclusters at different redshifts: a single fixed angular aperture will suffice.  However the size of the aperture could be adjusted to select different mass protoclusters.  Additionally, the large spatial extent of protoclusters means that using off-centre galaxies as a field sample may not produce a clean sample. As emphasised in Figure \ref{fig:ext}, the complex structure of protoclusters means they can be very extended in one direction. Therefore, even at large radii from the protocluster core there may be dense regions of protocluster galaxies.  To be certain of having a clean field sample for comparisons, field galaxies should be selected from regions more than 20 arcmin from the protocluster core. This conclusion correlates with low redshift theoretical and observational studies of clusters that have also emphasised the importance of selecting field samples far away from the cluster \citep[e.g.][]{Bahe13,Haines15}.

The next generation of large galaxy surveys, such as the Large Synoptic Survey Telescope (LSST) and \textit{Euclid}, have the potential to locate many protoclusters and high redshift clusters.  Existing techniques used to detect clusters at low redshift, such as X-ray identification or using the red sequence, are not suitable for protoclusters as they are not evolved enough to possess these properties.  However, one method of detecting protoclusters that would be suitable is to use an environment measure to identify overdensities.  In Figure \ref{fig:env} we plot the the third nearest neighbour density against galaxy stellar mass for all galaxies in the Millennium Simulation and those we have defined as protocluster members (any galaxy that will merge into the $z=0$ clusters).  The environment is characterised using $\delta$:

\begin{figure}
  \psfrag{x}[][][1][0]{${\rm log}[M_{*}/(h^{-1}{\rm M_{\odot}})]$}
  \psfrag{y}[][][1][0]{${\rm log}(1+\delta)$}
  \psfrag{a}[l][][1][0]{${\rm All~Galaxies}$}
  \psfrag{b}[l][][1][0]{${\rm Protocluster~Galaxies}$}
  \includegraphics[width=86mm]{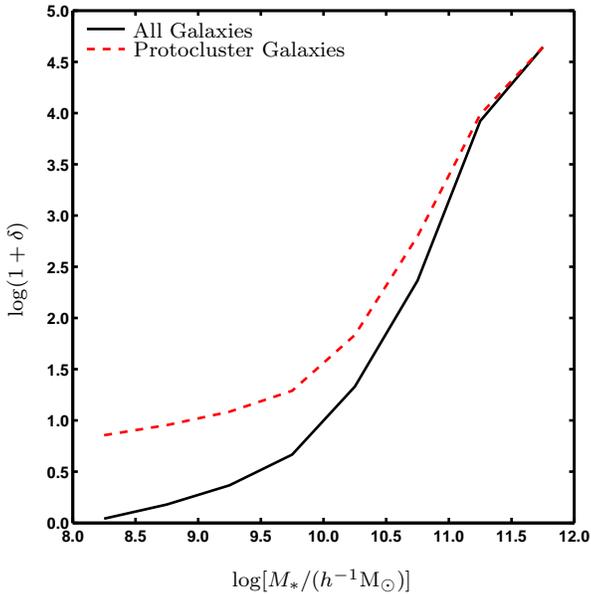}
  \caption{The 3rd nearest neighbour galaxy density of protocluster galaxies (red dashed line) relative to all galaxies (solid black line), as a function of stellar mass.  High mass galaxies show similar environments due to most of them being in protoclusters, however low mass galaxies diverge.}
\label{fig:env}
\end{figure}

\begin{figure}
\begin{center}
  \psfrag{x}[][][1][0]{${\rm log}[M_{*}/(h^{-1}{\rm M_{\odot}})]$}
  \psfrag{y}[][][1][0]{${\rm Purity}$}
  \includegraphics[width=86mm]{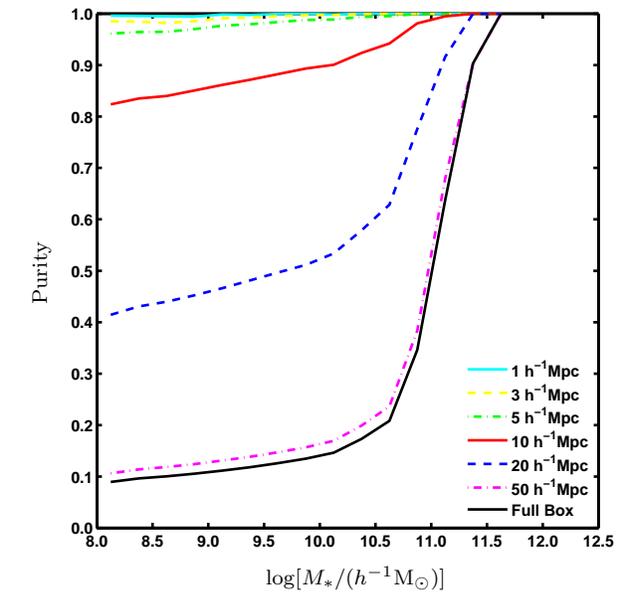}
  \caption{The fraction of $z=2$ galaxies within a given comoving volume, centred on the largest protocluster halo, that are protocluster members. Apertures are defined as the side length of a cube.  Small apertures that are typical of H$\alpha$ narrow-band imaging produce low contamination.}
\label{fig:frac}
\end{center}
\end{figure}

\begin{equation}
\delta=\frac{\rho-\bar{\rho}}{\bar{\rho}}=\frac{\rho}{\bar{\rho}}-1
\end{equation}

\noindent where $\rho$ is the galaxy density and $\bar{\rho}$ is the average density of all galaxies.  As expected, there is a clear trend for massive galaxies to reside in denser environments.  For very massive galaxies there is little difference between the environments occupied by protocluster galaxies and all galaxies.  This implies that most massive galaxies at high redshift reside in protoclusters.  For lower mass galaxies, the two curves diverge showing that the 3rd nearest neighbour density measure can pick out the protocluster overdensity relative to the field for all masses.  This means that measuring the environment of low mass galaxies around high mass galaxies offers the opportunity to locate protoclusters in large photometric redshift surveys.  Measuring accurate environments is more difficult at high redshift \citep{Shattow13} and the ability to accurately detect protoclusters using this method will be explored in future work.

Finally, we look at the level of contamination associated with the size of the aperture.  As we have seen, the limitations due to instrumental fields-of-view mean that we are limited to small apertures which do not capture the full protocluster.  Using a larger aperture would capture the whole protocluster, but would introduce a higher level of contamination from non-protocluster members.

In Figure \ref{fig:frac} we plot the fraction of galaxies that are protocluster members for different masses within different sized apertures.  Apertures are defined as the side length of a cube.  For all apertures there is little contamination at the very high mass end reaffirming our conclusion that most high mass galaxies are in protoclusters. This also reaffirms the previous conclusion that in order to attain a clean sample of field galaxies it is important to search further than 20 arcmin from the protocluster, as many galaxies, especially those with $M_{*}>10^{10.5}{\rm M_{\odot}}$, closer than this are likely to be protocluster members.  For small apertures the level of contamination is low at all masses reflecting the fact that only a small amount of the protocluster is being detected.  For large apertures the contamination is significant for low mass galaxies and by $50\,h^{-1}{\rm Mpc}$ it is the same as randomly sampling the Universe.  This implies that small apertures ($<10\,h^{-1}{\rm Mpc}$) are required to produce a clean sample.  If the protocluster galaxies are defined as all those that enter the virial radius by $z=0$, all of these purity fractions decrease by approximately a few to 10 per cent.  Furthermore, the contamination levels presented here are lower limits due to the use of cubes.  In reality the $z$ dimension would be significantly larger than the other two dimensions, in most cases, due to redshift uncertainties.  This would increase the level of contamination.

\begin{figure}
  \psfrag{a}[][][1][0]{$z$}
  \psfrag{b}[][][1][0]{$M/M_{200}^{z=0}$}
  \psfrag{c}[l][][1][0]{$1\le M_{200}^{z=0}<4\times10^{14}\,h^{-1}{\rm M_{\odot}}$}
  \psfrag{d}[l][][1][0]{$4\le M_{200}^{z=0}<10\times10^{14}\,h^{-1}{\rm M_{\odot}}$}
  \psfrag{e}[l][][1][0]{$M_{200}^{z=0}\ge 10^{15}\,h^{-1}{\rm M_{\odot}}$}
  \includegraphics[width=86mm]{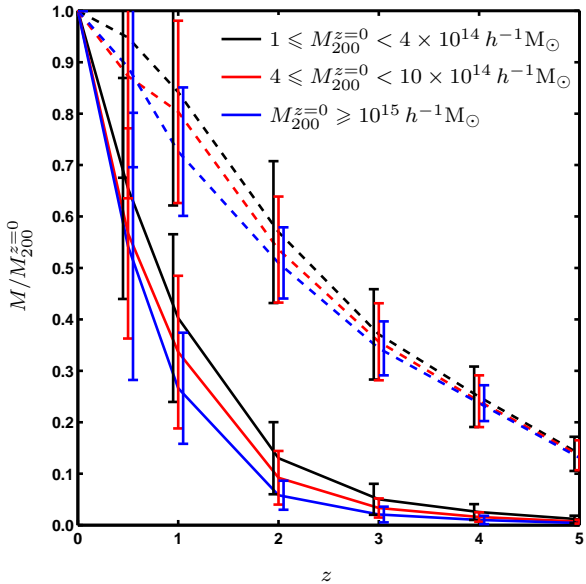}
  \caption{The evolution of halo mass for clusters binned by $z=0$ mass.  The lines shows the fraction of mass in the main halo (solid line) and all haloes that will merge to make the final cluster (dashed line) relative to the $z=0$ mass of the halo for different $z=0$ mass clusters.  Most of the protocluster mass is spread amongst many haloes and not concentrated in the main halo.  Error bars represent $1\,\sigma$ scatter and are offset about the middle mass bin by $\delta z=0.05$ for clarity.}
\label{fig:evo}
\end{figure}

\begin{figure}
  \psfrag{a}[][][1][0]{$z$}
  \psfrag{b}[][][1][0]{$N_{\rm MH}/N_{\rm Proto}$}
  \psfrag{c}[l][][1][0]{$1\le M_{200}^{z=0}<4\times10^{14}\,h^{-1}{\rm M_{\odot}}$}
  \psfrag{d}[l][][1][0]{$4\le M_{200}^{z=0}<10\times10^{14}\,h^{-1}{\rm M_{\odot}}$}
  \psfrag{e}[l][][1][0]{$M_{200}^{z=0}\ge 10^{15}\,h^{-1}{\rm M_{\odot}}$}
  \psfrag{f}[l][][1][0]{$M_{*}>10^9\,h^{-1}{\rm M_{\odot}}$}
  \psfrag{g}[l][][1][0]{$M_{*}>10^{10}\,h^{-1}{\rm M_{\odot}}$}
  \includegraphics[width=86mm]{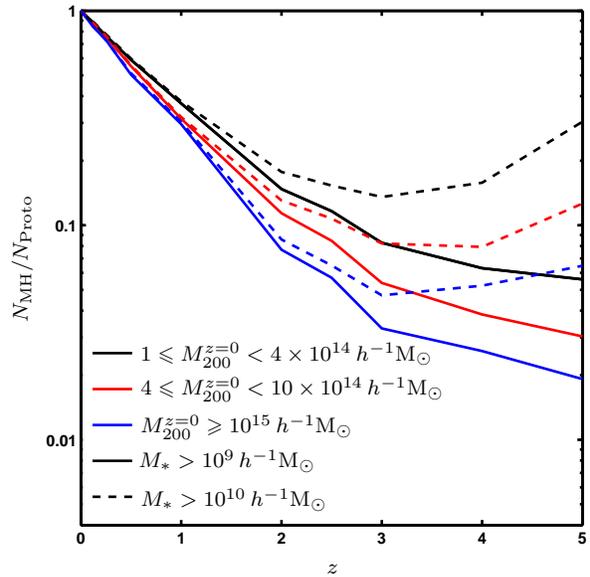}
  \caption{The fraction of galaxies in the main halo compared to the full protocluster with redshift.  Solid lines are for galaxies with stellar mass $M_{*}>10^9\,h^{-1}{\rm M_{\odot}}$, while dashed lines are for $M_{*}>10^{10}\,h^{-1}{\rm M_{\odot}}$.  Colours correspond to different cluster masses binned by $z=0$ mass.  Galaxies of higher mass are more likely to be found in the main halo.}
\label{fig:evogal}
\end{figure}

\subsection{The relationship between a protocluster and its main halo}
\label{sec:MainHalo}

In Figure \ref{fig:ext} we showed that the protocluster environment is a complex ``clumpy'' structure.  We define the main halo as being the most massive progenitor halo in the protocluster at a given redshift. If the main halo is massive enough it will be observed as a high redshift cluster. 

\subsubsection{Main halo versus protocluster mass}

To explore how the growth of the main halo relates to the protocluster as a whole, in Figure \ref{fig:evo} we plot the mass assembly of both.  The solid lines correspond to the fraction of the $z=0$ halo mass that is present in the largest progenitor halo with redshift.  The dashed lines correspond to the fraction of mass in all haloes hosting a galaxy of at least $M_{*}=10^8\,h^{-1}{\rm M_{\odot}}$, at that redshift, that will merge to form the final cluster.  The assembly plot has been grouped by final mass and, as expected from hierarchical growth, high mass clusters build up their mass later, although this is within the $1\,\sigma$ error.  

The rate of mass growth of the main halo differs from that of the whole protocluster. The growth of the main halo is slower than the rest of the haloes in the protocluster at $z>2$, but at $z<2$ the main halo grows more rapidly. This is a manifestation of the hierarchical growth of dark matter structures. In the early Universe there is a rapid increase in the number of small dark matter haloes in the protocluster that become large enough to host a galaxy. Also, at low redshift ($z<0.5$), there are significantly fewer dark matter haloes remaining in the protocluster which have not yet merged with the main halo. 

Figure \ref{fig:evo} illustrates that less than $\sim 20~{\rm per~cent}$ of the protoclusters mass is in the main halo at high redshift ($z>2$).  For $z=2$ the main halo contains only $\sim10$ per cent of the $z=0$ cluster's mass for a massive halo.  This means that studying only the main progenitor ignores a vast amount of information about the forming cluster.

\begin{figure}
  \psfrag{a}[][][1][0]{${\rm log}[M_{200}^{z=0}/(h^{-1}{\rm M_{\odot}})]$}
  \psfrag{b}[][][1][0]{${\rm log}[M_1/(h^{-1}{\rm M_{\odot}})]$}
    \psfrag{c}[][][1][0]{$M_2/M_1$}
  \includegraphics[width=86mm]{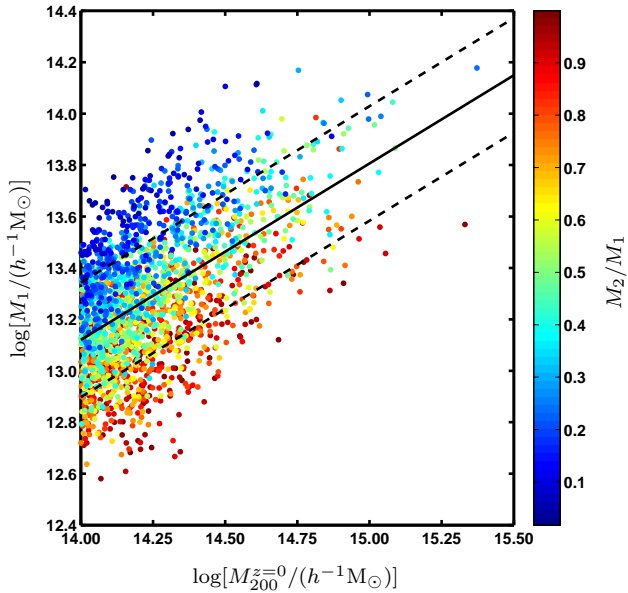}
  \caption{The mass of the most massive progenitor halo ($M_1$) at $z=2$ for a given $z=0$ cluster mass.  Points are coloured by the ratio of the mass of second most massive progenitor ($M_2$) to the first most massive ($M_1$) at $z=2$. The solid black line gives the best fit to the data with dashed lines representing the $1\,\sigma$ scatter about this.}
  \label{fig:Mrat}
  \end{figure}

Figure \ref{fig:evo} considered the growth of the dark matter halo.  We now look at the galaxies that reside in the main halo compared to the protocluster by plotting the fraction of galaxies in the main halo in Figure \ref{fig:evogal}.  These fractions are determined using two different stellar mass cuts.  As the redshift decreases the fraction of protocluster galaxies in the main halo increases for $M_{*}>10^9\,h^{-1}{\rm M_{\odot}}$.  This is expected as merging with the main halo brings galaxies that were residing outside of it in.  

For a higher stellar mass cut of $M_{*}>10^{10}\,h^{-1}{\rm M_{\odot}}$, however, there is an unexpectedly different trend.  As the redshift decreases to $z=3$, the fraction of galaxies in the main halo decreases before increasing again after this point.  Massive galaxies cannot leave the main halo, therefore this effect can only occur by galaxies outside of the main halo gaining enough mass to enter the sample.  These trends are apparent for all mass protoclusters.  Additionally, the most massive halo of higher mass protoclusters is less significant as it hosts a smaller fraction of the total galaxies.  

\begin{figure}
  \psfrag{a}[][][1][0]{${\rm log}[M_{200}^{z=0}/(h^{-1}{\rm M_{\odot}})]$}
  \psfrag{b}[][][1][0]{${\rm log}[M_1/(h^{-1}{\rm M_{\odot}})]$}
  \includegraphics[width=86mm]{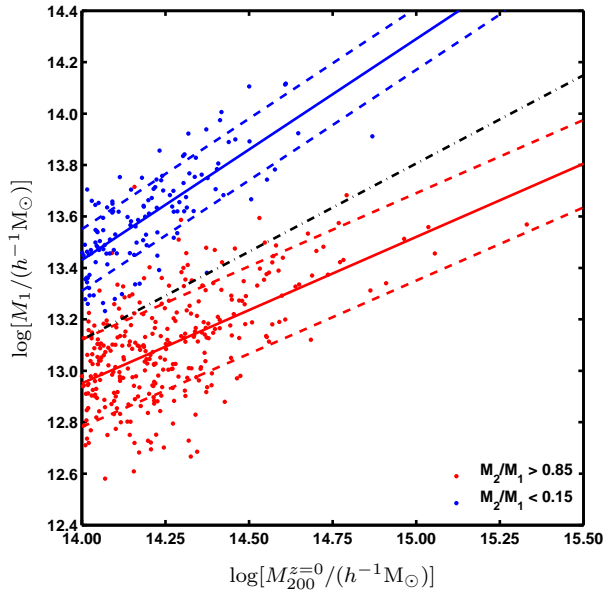}
  \caption{Same as Figure \ref{fig:Mrat}, but this time only protoclusters with a mass ratio of the second most massive progenitor to the first most massive ($M_2/M_1$), at $z=2$, greater than 0.85 (red) and less than 0.15 (blue) are shown.  The solid lines give the best fit to the data with dashed lines representing the $1\,\sigma$ scatter about this.  The black dot-dashed line represents the fit to all $M_2/M_1$ values.}
  \label{fig:bt}
  \end{figure}

Regardless of cluster mass, no main halo hosts more than 30 per cent of protocluster galaxies with $M_{*}>10^9\,h^{-1}{\rm M_{\odot}}$ at $z>2$. Observations that only study galaxies within the main progenitor halo (i.e.~the high redshift cluster) miss the majority of cluster galaxy progenitors. To trace the evolution of cluster galaxies it is essential that a representative fraction of the protocluster is observed.

\subsubsection{Estimating the $z=0$ cluster mass from the main halo mass at high redshift} 

\begin{figure*}
\begin{center}
  \psfrag{f}[][][1][0]{${\rm log}[M_{200}^{z=0}/(h^{-1}{\rm M_{\odot}})]$}
  \psfrag{d}[][][1][0]{${\rm log}[M_{\rm prog}/(h^{-1}{\rm M_{\odot}})]$}
    \psfrag{g}[][][1][0]{${\rm fraction}$}
     \psfrag{e}[][][1][0]{${\rm fraction}$}
  \includegraphics[width=86mm]{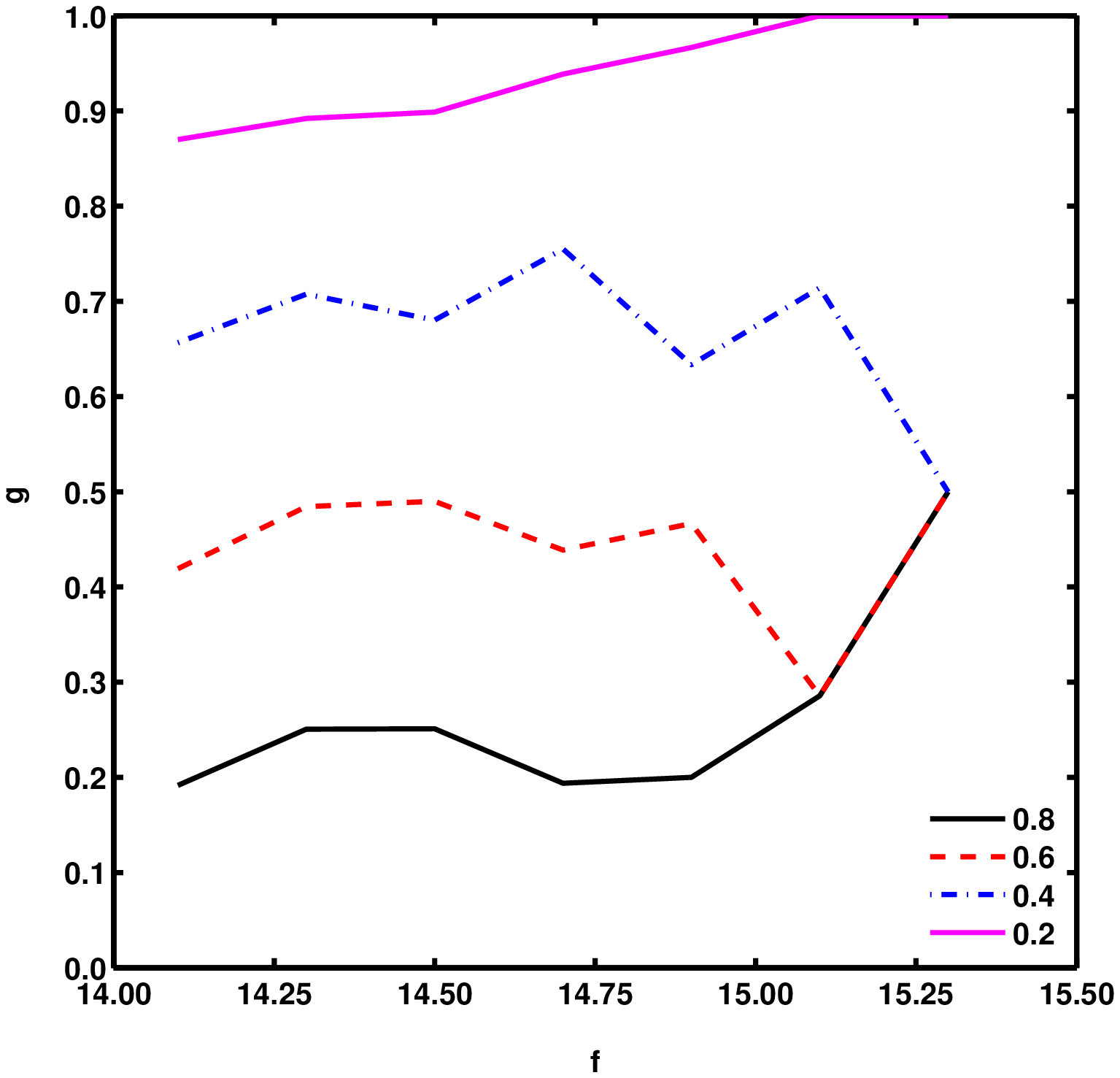}
   \includegraphics[width=86mm]{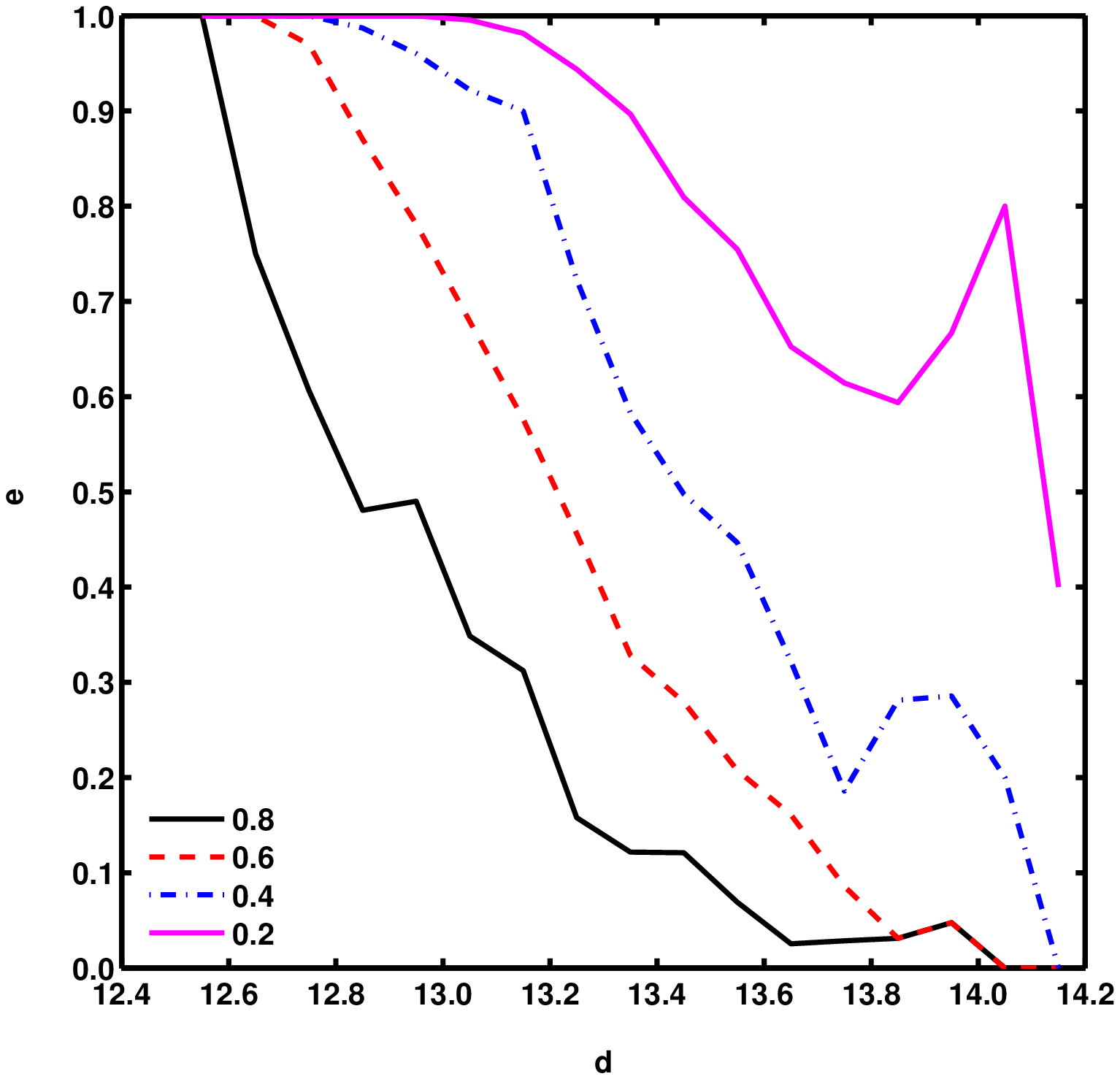}
  \caption{The fraction of protoclusters at $z=2$ where the second most massive halo is greater than 0.8 (black solid), 0.6 (red dashed), 0.4 (blue dot-dashed) or 0.2 (magenta solid) times the mass of the most massive halo.  The left panel presents this against the $z=0$ cluster mass while the right panel is against the most massive halo in the protocluster. }
\label{fig:Maj}
\end{center}
\end{figure*}

Having shown that the main halo contains only a fraction of the mass and galaxies of the protocluster, we now explore how much the structure of the protocluster can reveal about its evolutionary state, and the mass of the cluster it will become by $z=0$. The evolutionary state of a protocluster is described by the fraction of matter already located within the main protocluster halo. Protoclusters that contain a high fraction of their mass within the main halo are defined as further evolved. 

Figure \ref{fig:Mrat} plots the mass of the main halo at $z=2$ against the mass of the cluster at $z=0$. The median best fit to the data and $1\,\sigma$ deviation are shown by the black solid and dashed lines. These are determined by bootstrap sampling 100,000 times the least-squares fit.  There is a clear correlation between the mass of the main progenitor halo at $z=2$ and the mass of the resultant $z=0$ cluster: more massive progenitors tend to evolve into more massive clusters. However, for a given $z=0$ cluster mass, there is a large scatter in the range of masses for the main halo of the protocluster at $z=2$. This means that protoclusters exist in a range of evolutionary states at high redshift, which is nearly independent of the mass they will grow to by $z=0$. Thus estimating the $z=0$ mass of a cluster by extrapolating the mass of the high redshift cluster should be considered highly uncertain due to variation in the accretion history during cluster formation. 

If more than one progenitor halo can be identified, and its mass measured, the accuracy and precision of the extrapolation will increase.  Each point in Figure \ref{fig:Mrat} is colour coded to indicate the ratio of the mass of the second most massive halo in the protocluster ($M_2$) to the most massive ($M_1$) at $z = 2$, i.e. an indication of the dominance of the main halo in the protocluster. Clusters of all mass have a huge range in this ratio at higher redshift indicating the stochastic nature of cluster formation. The scatter in the relation between the cluster's mass at $z=2$ and $z=0$ separates into clear bands of protoclusters at different evolutionary states at that redshift. Therefore the mass ratio of the two most massive protocluster haloes ($M_1$ and $M_2$) provides an approximation of the evolutionary state of the protocluster. 

Figure \ref{fig:bt} shows two of these bands: protoclusters with dominant main haloes (blue points; $M_2/M_1<0.15$) and protoclusters in which there is no single dominant halo (red points; $M_2/M_1>0.85$). The solid lines represent the median best fit to these data, with dashed lines showing the $1\,\sigma$ scatter about these lines.  Both are again obtained by bootstrap sampling 100,000 times the least-squares fit. Clusters with a single, dominant progenitor halo (blue points) tend to have larger $z=2$ masses than those with higher $M_2/M_1$ ratios (red points) and correlate more strongly with $z=0$ cluster mass. 

Extrapolating the main halo mass to the $z=0$ cluster mass, whilst taking into account the mass ratio of the two most massive haloes within the protocluster, will not only improve the precision but also the accuracy.  For protoclusters with $M_2/M_1<0.15$, the scatter in the mass of the main halo at $z=2$ for a given $z=0$ cluster mass halves.  Additionally, the accuracy of this measurement increases, which can be quantified by considering the RMS of the difference between the predicted and true $z=0$ mass cluster.  For protoclusters with $M_2/M_1<0.15$, the estimate of the $z=0$ mass made without any information about the $M_2/M_1$ ratio (i.e. using the black dot-dashed fit) would have a 0.54 dex RMS deviation from the true mass, however this decreases to 0.15 dex when the ratio is taken into account (using the blue fit).  Observational studies which wish to estimate the $z=0$ mass of a protocluster may therefore improve the accuracy and precision of their estimate by measuring the mass of the two most massive haloes in the protocluster.

The masses of the two most massive haloes within a protocluster can be measured observationally in a number of ways. Galaxy velocity dispersions can give an estimate of the dynamical mass under the assumption that the galaxies are in virial equilibrium. Such a method has been used by \citet{Shimakawa14} to estimate the mass of two groups in a $z=2.53$ protocluster, finding a ratio of $\sim0.1$.  Alternatively, the mass of the haloes can be measured through observations of the intracluster gas using sensitive X-ray observatories, or at submilimetre wavelengths to detect the Sunyaev-Zel'dovich (SZ) decrement.  Current instrumentation is only able to measure collapsed structures at $z>1.5$ with masses greater than $10^{13.7}\,{\rm M_{\odot}}$ \citep[e.g.][]{Stanford12,Brodwin12,Andreon14}, but forthcoming instrumentation, such as the ESA \textit{Athena} satellite and the full ALMA array, will be able to detect much smaller groups at high redshift.  If the ratio of $M_2/M_1$ is sufficiently small, then it may not be possible to measure the mass of $M_2$ directly through X-rays or the SZ decrement. In this case it may be sufficient to measure the mass of $M_1$ through a detection of the intracluster medium and estimate the mass of $M_2$ through the ratio of stellar mass enclosed in $M_2$ and $M_1$. The stellar mass is a good tracer of the total cluster mass at low and intermediate redshifts (\citealt{Mulroy14}, Ziparo et al. in prep.) and the next generation of accurate cosmological simulations will help determine the relationship between the stellar mass and the total mass within more distant groups. 

\subsubsection{The importance of the main halo within a protocluster} 

\begin{figure}
  \psfrag{a}[][][1][0]{$n{\rm th~Massive~Galaxy}$}
  \psfrag{b}[][][1][0]{$M_{*enc}^{n}/M_{*enc}^{n=1}$}
   \psfrag{c}[l][][0.95][0]{$1.1\,h^{-1}{\rm Mpc~(1~arcmin)}$}
    \psfrag{d}[l][][0.95][0]{$2.8\,h^{-1}{\rm Mpc~(2.5~arcmin)}$}
    \psfrag{e}[l][][0.95][0]{$7.7\,h^{-1}{\rm Mpc~(7~arcmin)}$}
  \includegraphics[width=86mm]{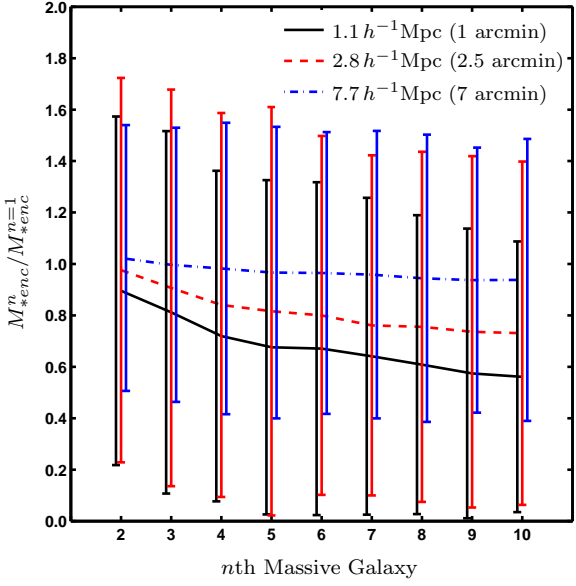}
  \caption{The effect of centring on the stellar mass enclosed in a cubical aperture of side length $1.1\,h^{-1}{\rm Mpc}$ (solid line), $2.8\,h^{-1}{\rm Mpc}$ (dashed) and $7.7\,h^{-1}{\rm Mpc}$ (dot-dashed).  Each curve gives the total stellar mass enclosed when the $n$th most massive galaxy in the protocluster is chosen as the centre with respect to the most massive.  For large apertures the difference in stellar mass is small however the scatter between protoclusters is large.  Error bars represent the $1\, \sigma$ scatter and are offset for clarity.}
\label{fig:centre_mass}
\end{figure}

We further explore the scatter in cluster formation history by showing the fraction of protoclusters where the mass ratio of the second to the first most massive halo is more than a given value.  The left panel of Figure \ref{fig:Maj} shows that the second most massive halo is at least $80~{\rm per~cent}$ of the mass of the most massive halo in $\sim 20{\rm ~per~cent}$ of protoclusters.  Only 10 per cent of protoclusters have a dominant main halo where the next largest is less than 20 per cent. These results are independent of the $z=0$ mass of the cluster except in the largest mass bin where there are few objects, which means the accretion history of clusters is erratic for clusters of all mass. The right panel of Figure \ref{fig:Maj} shows the same mass ratios, but this time against the mass of the main halo at $z=2$.  A clear mass dependence can be seen with large main haloes significantly less likely to have a massive companion in the protocluster.  

Overall, these Figures illustrate that the largest halo in many protoclusters should not be considered the dominant halo as it is often not significantly larger than other haloes.   These results also suggest that the observed examples of main halo dominated protoclusters is the result of potential bias in the observations.  Larger main haloes are easier to detect using current cluster-finding techniques (e.g.~red sequence algorithms, X-ray or SZ detections). Since it is easier to locate high redshift clusters with massive first ranked haloes, this subsample of evolved protoclusters will dominate observations. However, a significant fraction of massive clusters at $z=0$, do not yet have massive dominant haloes during the protocluster stage at high redshift. These less-evolved cluster progenitors would be missed by surveys searching for high redshift clusters, which target the most massive objects, and yet, they are equally likely to evolve into massive clusters by $z=0$. It is possible that the accretion history of a cluster leaves a lasting trace on its gas properties and the distribution of its galaxies. To trace the different evolutionary paths taken by collapsing clusters we must search for all types of cluster progenitors. 

\begin{figure}
  \psfrag{x}[][][1][0]{${\rm log}[M_{*}/(h^{-1}{\rm M_{\odot}})]$}
  \psfrag{y}[][][1][0]{${\rm log}[\Phi/(h^{3}{\rm Mpc^{-3}}{\rm dlog(M)})]$}
  \includegraphics[width=86mm]{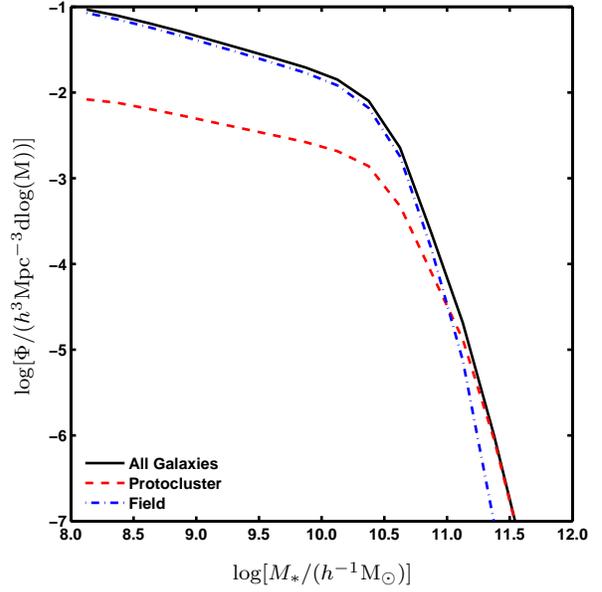}
  \caption{The galaxy stellar mass function for all galaxies within the simulation at $z=2$ (black solid line), those tagged as protocluster galaxies (red dashed line) and those tagged as field galaxies (i.e. not protoclusters; blue dot-dashed line).  The protocluster mass function has more massive galaxies and a shallower low mass slope.}
\label{fig:mfunc}
\end{figure}

If the whole of the protocluster cannot be viewed, and it is not clear if there is a main halo, it is important to see if the choice of the image centre has an effect on the results obtained.  For protoclusters selected due to the presence of a radio-loud AGN \citep[such as the Clusters Around Radio-Loud AGN (CARLA) Survey;][]{Wylezalek13}, often the centre of the protocluster is assumed to be the position of the radio-loud AGN, which is typically one of the most massive galaxies in the protocluster.  

In Figure \ref{fig:centre_mass} we present the stellar mass enclosed by various sized apertures, centred on the 10 most massive galaxies in the protocluster at $z=2$, with respect to that centred on the most massive.  For an aperture of $7.7\,h^{-1}{\rm Mpc}$ (7 arcmin) the difference in enclosed mass between an aperture centred on the 1st and 10th most massive galaxy is small, dropping to 94 per cent of the 1st most massive.  For the smallest aperture tested of $1.1\,h^{-1}{\rm Mpc}$ (1 arcmin), however, the drop is significantly larger, decreasing to 56 per cent mass of the one centred on the most massive galaxy.  An analogous calculation can be made for enclosed star formation rate which shows a smaller change.  

The most noticeable part of Figure \ref{fig:centre_mass} is the huge scatter between protoclusters.  For the smallest aperture tested, the mass enclosed by an aperture centred on the 10th most massive galaxy can vary from a decrease by a factor of 0.03 to an increase by a factor of 5 on that of an aperture centred on the most massive galaxy.  This large variation means there is a large uncertainty associated with the observed mass and star formation rate of protoclusters if the main halo cannot be readily identified. Observing large samples of protoclusters removes much of this uncertainty, therefore large statistical studies of protoclusters are less affected by this issue. 

\subsection{Biases introduced by observing only star-forming protocluster galaxies}
\label{sec:LowMass}

\begin{figure}
  \psfrag{x}[][][1][0]{${\rm log}[M_{*}/(h^{-1}{\rm M_{\odot}})]$}
  \psfrag{y}[][][1][0]{${\rm SF~Fraction}$}
  \includegraphics[width=86mm]{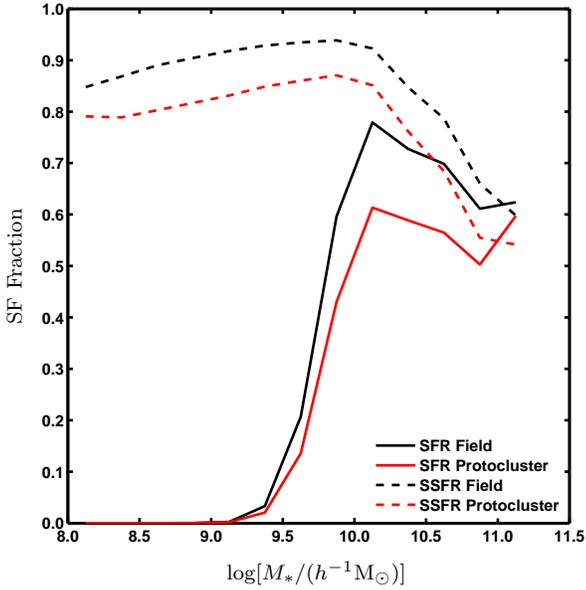}
  \caption{The fraction of star forming galaxies as a function of mass using two different star formation cuts.  The solid lines correspond to a fixed cut in SFR \citep{Cooke14}, while the dashed line corresponds to a fixed cut in sSFR \citep{Lani13}.  A fixed SFR cut, such as that in H$\alpha$ narrow-band observations, can cause the artificial loss of star forming low mass galaxies.}
\label{fig:star}
\end{figure} 

A very common technique to locate a clean sample of protocluster galaxies is to select line-emitting galaxies, such as H$\alpha$, Ly$\alpha$ and [O{\sc ii}] emitters, using narrow filters  \citep[e.g.][]{Venemans07,Koyama13}. However, this method is only able to identify the active subset of protocluster galaxies. Here we explore how our interpretation of the galaxy stellar mass function and overdensity of the protocluster can be affected by only studying the star forming galaxy population. 

In Figure \ref{fig:mfunc} we plot the galaxy stellar mass function for protocluster members (red dashed line), field galaxies (non-protocluster members; blue dot-dashed line) and all galaxies (the sum of the two; black solid line) at $z=2$.  The shape of the protocluster mass function compared to the field differs slightly.  There are more massive galaxies in protoclusters than the field, despite there being more galaxies in general in the field.  
At the low mass end, the slope of the protocluster mass function is shallower than that of the field.  The value for the turnover ($M^*$) is fractionally higher for the protocluster.  The shallower low mass slope in the semi-analytic model protoclusters reduces the number of expected low mass galaxies compared to the field, but not greatly.

Conventional definitions of star forming and non-star forming galaxies involve cuts in Specific Star Formation Rate (sSFR; ${\rm SFR}/M_{*}$).  One such definition, used in \citet{Lani13} for example, is to define a galaxy as non-star forming if its mass doubling time, calculated from its present star formation rate, is more than the age of the Universe.  Applying this to our simulated $z=2$ galaxy sample yields a star forming galaxy fraction corresponding to the dashed line in Figure \ref{fig:star}.  This demonstrates that there are fewer star forming galaxies in the protocluster than the field, but in general they follow the same trend with mass.

\begin{figure}
  \psfrag{a}[][][1][0]{${\rm log}[M_{*}/(h^{-1}{\rm M_{\odot}})]$}
  \psfrag{b}[][][1][0]{${\rm SF~Fraction}$}
  \includegraphics[width=86mm]{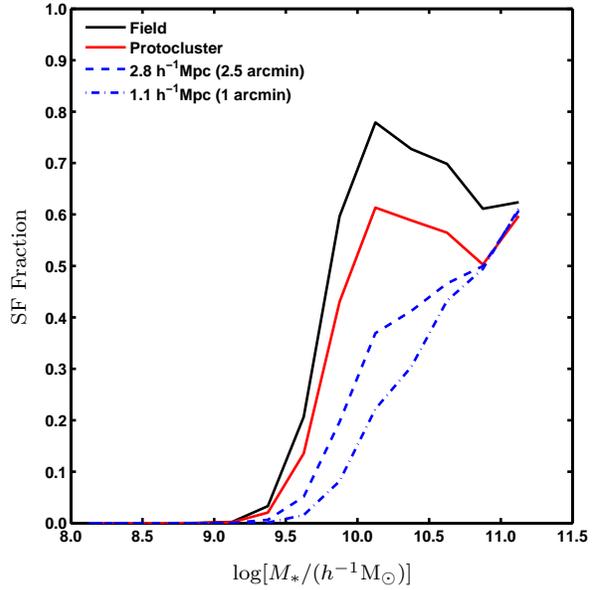}
  \caption{The fraction of star forming galaxies with respect to environment using a fixed SFR cut at $z=2$.  The black solid line corresponds to field galaxies, whilst the red solid line to protocluster galaxies.  The blue dashed and dot-dashed lines correspond to just galaxies within a $2.8\,h^{-1}{\rm Mpc}$ comoving (2.5 arcmin) and $1.1\,h^{-1}{\rm Mpc}$ comoving (1 arcmin) cube centred on the most massive protocluster galaxy respectively.  The larger of these is similar to the aperture used by \citet{Cooke14} and shows a strong environmental relation with the number of star-forming galaxies observed.}
\label{fig:app}
\end{figure}

Imaging in a narrow-band to detect emission line galaxies does not select galaxies based on their sSFR, but instead produces a cut in SFR.  This can have a different effect, especially at the low mass end, as a galaxy's SFR is dependent on its mass if it is on the main star forming sequence.  Applying a cut of $7\,{\rm M_{\odot}\,yr^{-1}}$ (typical of recent works, e.g. \citealt{Cooke14}), gives a star forming fraction that corresponds to the solid lines in Figure \ref{fig:star}.  This produces very different trend to that of the sSFR cut.  At low masses the star forming fraction rapidly descends to zero as the cut intercepts the star forming main sequence.  This leads to a minimum detectable mass for emission-line selected galaxies such as those observed in H$\alpha$ narrow-band images. Thus selecting galaxies by their star formation rate biases against low-mass galaxies. 

The limitations of instrumental field of view mean that only a small fraction of the protocluster is typically observed.  If we consider only the central region of the protocluster, rather than all members, we get a further bias in the results.  Figure \ref{fig:app} replots the star forming fraction of protocluster and field galaxies with a $7\,{\rm M_{\odot}\,yr^{-1}}$ cut (solid lines), but only including the star forming galaxies within the central $2.8\,h^{-1}{\rm Mpc}$ comoving (2.5 arcmin; blue dashed line) and $1.1\,h^{-1}{\rm Mpc}$ comoving (1 arcmin; blue dot-dashed line) regions of the protocluster.  Using the small window only captures the densest region where quenching is efficient and has led to a much lower fraction of star forming galaxies relative to the field.  

By observing only the star forming galaxies in the main halo of the protocluster we obtain a very biased view of the mass function of protocluster galaxies. Having a fixed threshold intercepts the star forming main sequence resulting in the suppression of galaxies detected below $10^{10}\,{\rm M_{\odot}}$ and a near total loss of galaxies below $10^{9}\,{\rm M_{\odot}}$. In addition to the loss of galaxies because they drop below the star formation rate threshold, the small windows used for narrow-band observations results in a further loss.  Focussing on just the very centre, as opposed to the full protocluster, significantly increases the quenched fraction of galaxies.  This is because more environmental quenching occurs within the densest part of the protocluster.  While larger apertures would reveal more of the protocluster, it would also increase the level of contamination of non-protocluster members.  For small apertures the sample has little contamination, but by $10\,h^{-1}{\rm Mpc}$ at least 20 per cent of low mass galaxies are interlopers.

An important side effect of losing low mass galaxies in narrow-band observations is that the measured overdensities for protocluster will be highly uncertain.   If the full observed sample of galaxies is used, then the absence of low mass galaxies in the protocluster compared with the field will lead to the overdensity being underestimated.  Using a mass cut however is also problematic.  The simulations indicate that almost all very massive galaxies reside in protoclusters and this will lead to an unrepresentative field sample, leading to a very high overdensity estimate.  Quantifying the overdensity accurately is important for estimating the eventual mass of the protocluster using the \citet{Chiang13} method.  Due to the above reasons, it is not advisable to estimate the mass of a protocluster from the overdensity measured from the excess of emission line galaxies in a small field of view.

\section{Implications for observations}
\label{sec:sum}

We have explored the difference between protoclusters and high redshift clusters using a semi-analytic model applied to the Millennium Simulation.  Clusters were identified as $z=0$ haloes with masses greater than or equal to $10^{14}\,h^{-1}{\rm M_{\odot}}$.  All galaxies that will merge to make these clusters were tagged at higher redshift and classed as protocluster members. The most massive virialised dark matter halo in the protocluster is defined as the main halo, and would be observed as a high redshift cluster or group if it were massive enough.

We find that protoclusters are very extended, with 90 per cent of the mass spread over $\sim35\,h^{-1}{\rm Mpc}$ comoving at $z=2$  ($11\,h^{-1}{\rm Mpc}$ physical; 30 arcmin).  This is far larger than the typical targeted observations of protoclusters being currently conducted using line-emitting galaxies.  This implies that these studies of protoclusters and high redshift clusters are not imaging all of the protocluster, but instead are focussed on only a small part of  the structure.  

The protocluster structure comprises many haloes linked by filaments.  This has important consequences for the evolution of cluster galaxies, since not all galaxies that make up the cluster at $z=0$ have had the same environmental history.  Some will have formed in the main halo, others will have been residing in smaller haloes or in filaments for much of their history.  Thus the environmental history of cluster galaxies is complex and non-uniform. Some galaxies experience strong `environment preprocessing', where galaxies experience environmental effects prior to cluster infall, whereas others do not \citep[e.g.][]{DeLucia2012}. 

We find that the largest halo of the protocluster only hosts a minority of protocluster galaxies at high redshift, with typically less than 20 per cent of galaxies with $M_{*}>10^9\,h^{-1}{\rm M_{\odot}}$ residing within it at $z>2$. To study the evolution of cluster galaxies it is therefore essential that a representative fraction of the protocluster is observed, and not simply the minority of protocluster galaxies that reside within the high redshift cluster core.  Whilst this will improve our understanding of the role of preprocessing, it does come at the expense of sample purity.

We have shown that only a small subset of protoclusters evolve as a single main halo with significantly smaller objects merging onto it. Only 10 per cent of protoclusters at $z=2$ are dominated by a single halo, i.e. where no other member haloes in the protocluster have more than 20 per cent of the main halo's mass.  A fifth of protoclusters exhibit very little difference between the most massive and second-ranked halo as the mass ratio is $>0.8$. Whether a protocluster contains a dominant halo at high redshift does not depend on its $z=0$ mass, however, if the first-ranked halo is very massive (so it would be detected as a high redshift group or cluster), then it is likely to be a very dominant halo. Observational techniques that are predisposed to locate protoclusters based on the mass of their main halo (e.g.\,X-ray or SZ detection) are biased to select the subset of protoclusters with single dominant haloes, and therefore are likely to miss the majority of cluster progenitors with no dominant halo. 

Having many large haloes in the same protocluster will additionally have important consequences for cluster cosmology.  The close proximity of large haloes in protoclusters will make it difficult to separate them observationally.  This may result in haloes being classed as a single more massive object and hence discrepant with the output of dark matter simulations.  

For over a decade studies of protoclusters have used narrow filters to isolate and study star-forming protocluster galaxies. This technique is popular as it efficiently selects a relatively clean sample of protocluster galaxies.  However, several recent observational studies have shown that the stellar mass function of star-forming galaxies in protoclusters differs from that of the field \citep[][Husband et al. in prep]{Steidel05,Hatch11b,Koyama13, Cooke14}. This means that the mass function of star-forming galaxies in protoclusters is no longer a scaled version of the field, and hence implies that the bias of this population depends on environment. This has severe implications for measuring the mass overdensity: the measured galaxy overdensity may not be correctly converted to a mass overdensity.

The semi-analytic model we have investigated suggests the observed difference in the stellar mass functions is due to environmental quenching of low mass star-forming galaxies. This effect is exacerbated if the observations are concentrated on the main halo where environmental quenching is strongest (Figure 13).  Observations taken with larger fields-of-view (greater than 10 comoving Mpc) will not be strongly impacted by these environmental effects, and thus the biases of the field and protocluster emission line galaxies will be similar on large scales \citep[as shown by][]{Chiang13}.  However, if the model prescription of the quenching is too aggressive, the cause of the observed mass function divergence may extend beyond the main halo, and impact mass overdensities determined even from large apertures. Future observations of the star-forming galaxy mass function on larger scales is needed to test the environment quenching scenario.  In summary, the mass overdensity measured from the excess of emission line galaxies should be considered unreliable, especially in small apertures, and should not be used to estimate the $z = 0$ mass of the protocluster.

\section{Conclusions}

As highlighted at the start of this paper, the term protocluster is used to describe the progenitors of galaxy clusters, but differing definitions are used in the literature.  Protoclusters are diffuse collections of haloes, linked by filaments, that will merge to make up the final low redshift clusters. These structures are very extended, with 90 per cent of the mass spread over $\sim15-35\,h^{-1}{\rm Mpc}$ comoving at $z=2$, with the radial extent depending on the final mass of the cluster. High redshift clusters are the manifestations of massive main haloes within protoclusters.  However in most cases the largest halo of the protocluster only hosts a minority of protocluster galaxies at high redshift, so a representative fraction of the protocluster must be observed to study the evolution of cluster galaxies. 

Protoclusters exist in a range of evolutionary states at high redshift, independent of the mass they will evolve to by $z=0$. Here we define evolution by the amount of $z=0$ cluster mass in the main halo.  Only a small subset of protoclusters host a dominant main halo that would be identifiable as a high redshift cluster. The evolutionary state of a protocluster can be approximated from the mass ratio of the first and second ranked haloes in the protocluster. Furthermore, a more accurate estimate of the mass of the $z=0$ descendant cluster can be determined if both the main halo mass and the evolutionary state of the protocluster are known.   

Large observations spanning several arcmin are required to view all the different physical processes that affect galaxies within forming clusters. The assembly history of clusters is varied and we must examine protoclusters both with and without dominant main haloes to understand the numerous paths by which clusters of galaxies form.  Future large scale observations of protoclusters will offer the opportunity to better understand both cluster formation, and the importance of environment history in galaxy evolution.

\section*{Acknowledgments}

The authors wish to thank Caterina Lani and Andrew King for useful discussion.  SIM acknowledges the support of the STFC consolidated grant ST/K001000/1 to the astrophysics group at the University of Leicester.  NAH is supported by a STFC Rutherford Fellowship.  EAC acknowledges the support of the STFC.

This work made use of the ALICE High Performance Computing Facility at the University of Leicester and the DiRAC Complexity system, operated by the University of Leicester IT Services, which forms part of the STFC DiRAC HPC Facility (www.dirac.ac.uk ). This equipment is funded by BIS National E-Infrastructure capital grant ST/K000373/1 and STFC DiRAC Operations grant ST/K0003259/1. DiRAC is part of the National E-Infrastructure.

The Millennium Simulation used in this paper was carried out by the Virgo Supercomputing Consortium at the Computing Centre of the Max-Planck Society in Garching.  The halo merger trees used in the paper are publicly available through the GAVO interface, found at http://www.mpa-garching.mpg.de/millennium/.

\bibliographystyle{mn2e}
\bibliography{proto}

\bsp

\appendix

\section{Cluster Size Definition}
\label{ap}

\begin{figure*}
\begin{center}
  \psfrag{a}[][][1][0]{$z$}
  \psfrag{b}[][][1][0]{$r_{\rm comoving}/(h^{-1}{\rm Mpc)}$}
  \psfrag{c}[l][][0.78][0]{$1\le M_{200}^{z=0}<4\times10^{14}\,h^{-1}{\rm M_{\odot}}$}
  \psfrag{d}[l][][0.78][0]{$4\le M_{200}^{z=0}<10\times10^{14}\,h^{-1}{\rm M_{\odot}}$}
  \psfrag{e}[l][][0.78][0]{$M_{200}^{z=0}\ge 10^{15}\,h^{-1}{\rm M_{\odot}}$}
    \psfrag{f}[][][1][0]{$r_{\rm physical}/(h^{-1}{\rm Mpc)}$}
      \psfrag{g}[][][1][0]{$r_{\rm angular}/{\rm arcmin}$}
  \includegraphics[width=180mm]{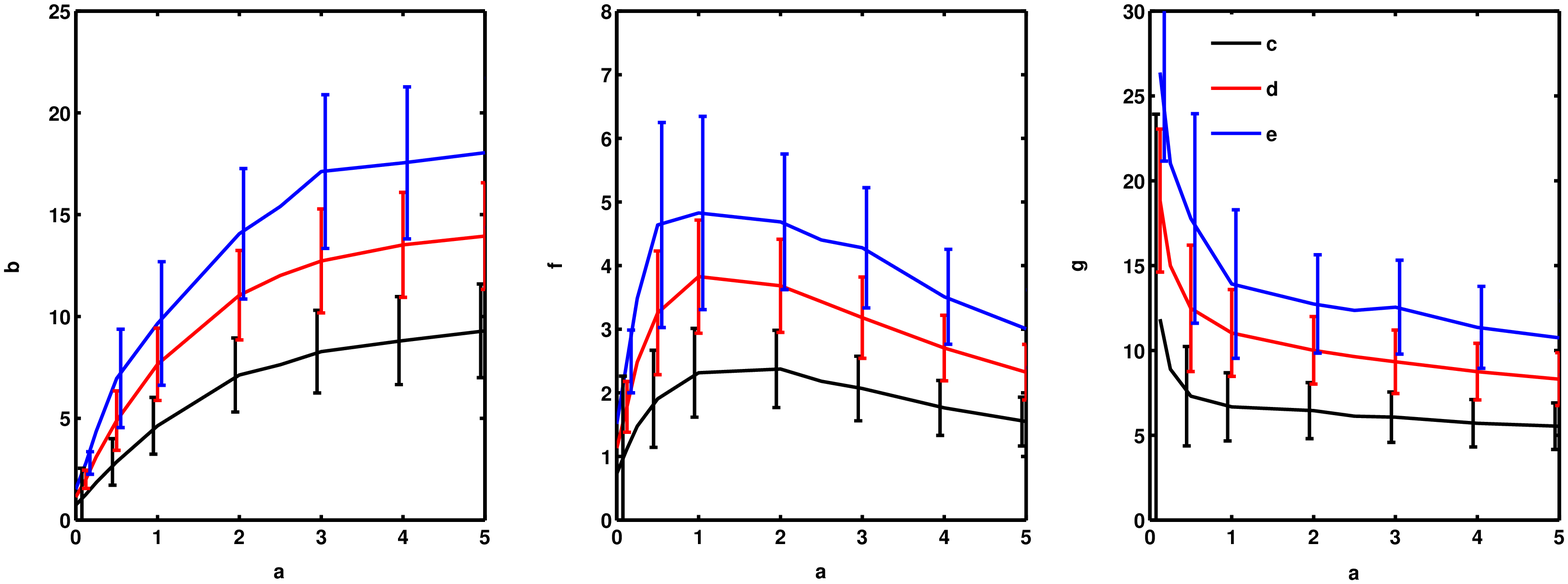}
  \caption{The average radius that encloses 90 per cent of the stellar mass of a protocluster at different redshift, for binned $z=0$ cluster masses. The left panel represents comoving radius, centre panel the physical radius and right panel the angular projection.  Error bars represent $1\,\sigma$ scatter and are offset about the middle mass bin by $\delta z=0.05$ for clarity.  Only galaxies that will be within the virial radius at $z=0$ are considered.}
  \label{fig:sizvir}
  \end{center}
\end{figure*}

Within this paper we have defined protocluster member galaxies as being any galaxy that merges and forms part of the friends-of-friends halo of a $z=0$ cluster.  An alternative definition would be to consider only those galaxies that reside within the virial radius of the cluster at $z=0$.  In Figure \ref{fig:sizvir} we reproduce Figure \ref{fig:rad}, this time using only galaxies that will be within the virial radius at $z=0$.  Using this definition results in smaller sizes, but the same evolutionary pattern is still present.  The choice of cluster definition will result in small changes to the absolute values quoted in this paper, but the overall results will remain the same.

\label{lastpage}

\end{document}